\documentclass{aa}
\begin{document}

   \thesaurus{03(02.01.2; 11.14.1)} 
\title{Semi-analytical radiative transfer in plane-parallel
geometry: application to accretion disk coronae}

   \author{Xinwu Cao\inst{1}, D.R. Jiang\inst{1}, J.H. You\inst{2},
   J.L. Zhao\inst{1}}

   \offprints{Xinwu Cao (cxw@center.shao.ac.cn)}
   
   \institute{ {\inst{1}
{Shanghai Observatory, Chinese Academy of Sciences, 80 Nandan Road,
Shanghai, 200030, P.R. China}}\\
{\inst{2}  {Department of Applied Physics, Shanghai Jiaotong University,
Shanghai, 200030, P.R. China}  }   }

   \date{Received  January , 1997; accepted August 28, 1997}
\authorrunning{X. Cao et al.}
\titlerunning{Semi-analytical radiative transfer in
plane-parallel geometry}

 \maketitle

   \begin{abstract}
A simplified frequency integrated 
radiative 
transfer 
equation is solved to study Compton 
scatterings in the corona of the
disk by using numerical iterating method. 
We find that the vertical
thickness of the corona cannot be used as the typical length to measure
the optical depth of the corona. A semi-analytical approach is  
proposed to calculate the energy dissipations in
the corona of the disk. We demonstrate that our approach can
reproduce the numerical solutions to an accuracy
of $<2\%$.

   \end{abstract}

\section{Introduction}

The so-called UV bump and the X-ray power law spectra are the most important
features in the spectra of Active Galactic Nuclei (AGN). It is well believed
that the luminosities of the AGN are produced by accretion of matter through
disks onto massive black holes.
Shields (1978) suggested that the UV bump observed in AGN could be
attributed to the thermal radiation from the disk. Malkan (1983) showed
that the UV bump in several quasars could be fitted with predicted spectra
of optically thick accretion disks.
The observed spectrum in the medium X-ray range is close to a power law,
with a small dispersion in the values of the spectral index, whose average
for Seyfert galaxies is $\sim$ 0.7 (Mushotzky 1984; Turner \& Pounds 1989).
In fact, the average X-ray spectral index becomes $\sim$ 0.9 after
subtracting the component of X-rays reflected by cold matter
(Pounds et al. 1990).

The standard geometrically thin, optically thick accretion disk model
(Shakura \& Sunyaev 1973) can be employed to explain the UV bump
(Malkan 1983), but the standard disk is too cold to produce the X-ray
radiation. Another solution for the inner
parts of accretion disks around black holes is the optically thin, hot
disk model, which was first proposed by Shapiro, Lightman \& Eardley (1976). Electron
temperature of the disk in this model becomes $\sim$ 10$^9$ K and it can
explain observed X-ray emissions. However, it fails to reproduce the
observed UV bump. Haardt \& Maraschi (1991) proposed a two-phase thermal disk-corona model.
In this model, most of the thermal soft photons emitted from the optically
thick, cold disk pass the optically thin corona without being scattered and
are observed
as UV bump. Only a small fraction of them is Compton upscattered to the
X-ray range by the high temperature thermal electrons in the corona. About
half of the Compton scattered photons are directed toward the disk and
are  reprocessed to emerge as blackbody radiation, while the remaining half
are directed upwards and are observed as power law X-ray spectrum.
In this model, the gravitational energy of the accreting matter is mainly
released in the corona of the disk. Along the line of this model, some workers investigated
the structure of the disk-corona system with different assumptions
(Nakumura \& Osaki 1993; Kusunose \& Mineshige 1994; Svensson \& Zdziarski
1994). The X-ray spectra emitted by the corona are studied in detail by using
different approaches, especially the Monte Carlo methods (Haardt 1993;
Haardt \& Maraschi 1993; Titarchuk 1994; Poutanen \& Svensson 1996).
Nakamura \& Osaki (1993) found in their numerical simulations that the
effective optical depth of photons traveling through the corona may be
larger than the optical depth in the normal direction of the disk.

In this work, we solve the integro-differential radiative transfer equation
by using numerical iterating method. The basic equations describing the
problem are listed in Sect. 2. Section 3 contains the results and a
semi-analytical approach proposed to reproduce the numerical results. The last
section is the discussion of the results.

\section{Basic equations   }

The frequency-integrated radiative transfer equation is

   \begin{equation}
  \mu{{\partial I(\tau, \mu)}\over {\partial \tau}}=I(\tau, \mu)-S(\tau),
  \end{equation}
where $\tau$ is the optical depth measured in the normal direction
of the disk, $I(\tau, \mu)$ and $S(\tau)$ are
the frequency-integrated specific intensity and source function,
$\mu=\cos\theta$, describes the direction of $I(\tau, \mu)$ with respect
to the normal direction.

Here we only consider pure Compton scatterings in the corona and assume
the Compton scatterings to be isotropic. Thus, the source function $S(\tau)$
could be written as

\begin{equation}
S(\tau)={{1+A}\over {4\pi}}\int I(\tau, \mu)d\Omega={{1+A}\over 2}
\int\limits_{-1}^{1}I(\tau, \mu)d\mu,
\end{equation}
where $A=4\Theta+16\Theta^2$, is the mean fraction  of the energy
change of the photon through single scattering. The dimensionless
electron temperature $\Theta=kT_{e}/m_{e}c^{2}$, where $k$, $m_e$ and
$c$ are the Boltzman constant, rest mass of the electron and the light speed.
Strictly speaking, the Compton scattering in the corona is anisotropic
(Ghisellini et al. 1991; Haardt 1993). Here we only intend to investigate the total energy
transferred from the mild relativistic electrons in the corona to the
scattered photons. In the first attempt, we employ Eq. (2) to describe the
source function in the radiative transfer equation for the sake of simplicity.
We believe this will not affect the main results of our present investigation.

We consider the homogeneous isothermal layer in a plane-parallel geometry,
which can model the hot corona over an accretion disk. Compton parameter $y$
is defined by

\begin{equation}
y=\tau_{0}(4\Theta+16\Theta^{2})\max (1, \tau_{0}), 
\end{equation}
where $\tau_{0}$ is the optical depth of the corona by electron scattering
measured in the vertical direction.
Now, we can solve Eq. (1) combining Eq. (2)
with appropriate boundary conditions, if the electron temperature
$\Theta$ is specified.

In this work, we assume the input soft photons from the cold disk are
isotropic, then the boundary conditions for the problem are as follows:

\begin{equation}
I(\tau, \mu)=I_{0},~~~~~\mu>0,~~~at~~~~~~ \tau=\tau_{0}  $$
\end{equation}
and
\begin{equation}
I(\tau, \mu)=0,~~~~~~~~~~\mu<0,~~~at~~~~~~ \tau=0, 
\end{equation}
where $\tau$ is the electron scattering optical depth measured in the normal 
direction of the disk. The energy flux of the soft photons from the cold disk is

\begin{equation}
F_{s}=2\pi\int\limits_{0}^{1} I(\tau_{0}, \mu)\mu d\mu=\pi I_{0}.
\end{equation}
Most of the soft photons emitted by the cold disk pass the optically
thin corona without being scattered. The flux of the soft photons escaped
from the system is given by

\begin{equation}
F_{esc}=2\pi\int\limits_{0}^{1} I(\tau_{0}, \mu)e^{-\tau_{0}/\mu}\mu
d\mu.
\end{equation}
A small fraction of soft photons from the cold disk is Compton upscattered
in the corona. About half of the scattered photons are directed upwards.
The remaining half are directed downwards, part of them are absorbed by the
cold disk and part are reflected.

The fluxes of the upward scattered photons and downward scattered photons
are

\begin{equation}
F_{uC}=2\pi\int\limits_{0}^{1} I(0, \mu)\mu d\mu-F_{esc}
\end{equation}
and

\begin{equation}
F_{dC}=2\pi\int\limits_{-1}^{0} I(\tau_{0}, \mu)\mu d\mu,
\end{equation}
respectively.

The fraction $D$ of the energy dissipated in the hot corona to the input
soft photon energy flux from the cold disk is then given by

\begin{equation}
D={ { F_{uC}+F_{dC}+F_{esc}-F_{s} } \over {F_{s}}  }.
\end{equation}
Finally, the fraction $\eta$ of Compton upscattered photon energy directed
downward to the cold disk is

\begin{equation}
\eta={ {F_{dC}} \over { F_{dC}+F_{uC} } }.
\end{equation}

\section{Results}

Equation (1) is an integro-differential equation and cannot be solved by simply
iterating, since the source function is not known a priori and depends
on the solution $I(\tau, \mu)$ at all directions through a given point.
In our numerical calculations, we first set the source function $S(\tau)=0$
and the test solution ${I_{1}}(\tau, \mu)$ is available. Let $I(\tau, \mu)
={I_{1}}(\tau, \mu)$, the source function ${S_{1}}$ is obtained from
Eq. (2). Then the more accurate solution  ${I_{2}}(\tau, \mu)$
could be obtained  by iterating Eq. (1), if we let $S(\tau)={S_{1}(\tau)}$
in Eq. (1). Similar to the first step we have taken, the solution ${I_{n}}
(\tau, \mu)$ and the source function $S_{n}(\tau)$ will be known by
repeatedly iteration. The iteration is terminated until it reaches
$({S_{n}}-{S_{n-1}})/S_{n}<10^{-5}$, then the solution to Eq. (1) is
available.

The numerical results on the fraction $D$ of the energy dissipated in the
hot corona to the input soft photon energy flux from the cold disk
are plotted in Fig. 1. It is found that the behaviour of $D$ varies with
the different electron temperatures $\Theta$.
For the higher electron
temperature, the value $D$ becomes larger even at the same Compton 
parameter $y$.

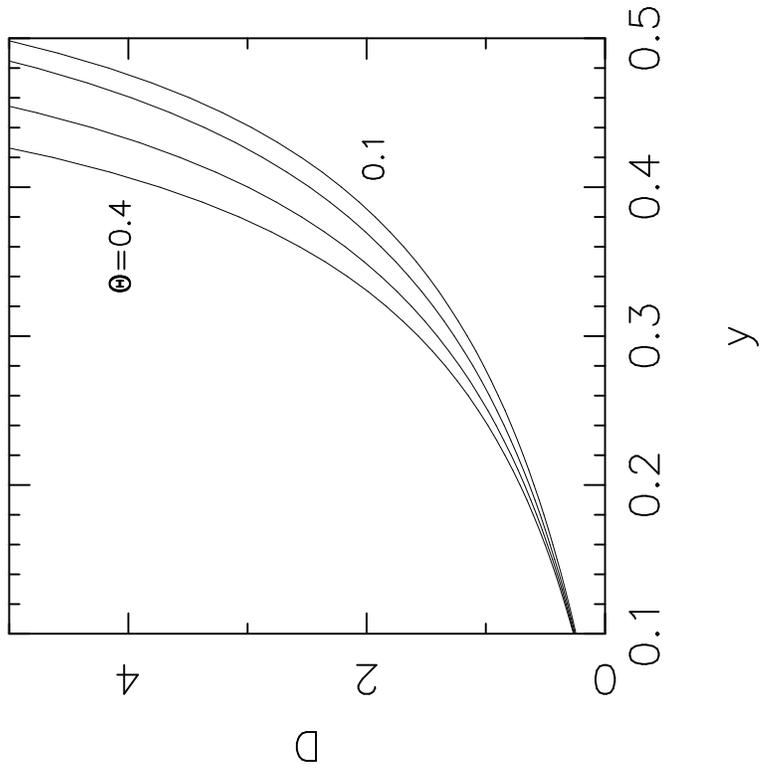
\begin{figure}
      \vspace{8cm}
     \caption{The numerical results of the fraction $D$ of the energy dissipated in
the corona to the input soft photon energy vs. Compton parameter $y$, with
respect to the electron temperatures $\Theta=0.1, 0.2, 0.3, 0.4$.}
\label{}
   \end{figure}

Considering that the soft photons only suffer single scattering in
the corona, we can obtain the mean energy gained by the photons from the
hot electrons in the optically thin corona is

\begin{equation}
\Delta E= A \int\limits_{0}^{1}I(\tau_{0},\mu)(1-e^{-\tau_{0}/\mu})
\mu d\mu=I_{0} y{{a(\tau_{0})}\over \tau_{0}},
\end{equation}
where

\begin{equation}
a(\tau_{0})=\int\limits_{0}^{1}(1-e^{-\tau_{0}/\mu})\mu d\mu,
\end{equation}
is the fraction of the photons which suffer single scattering in the
corona to the injecting isotropic soft photons.
It can be seen in Fig. 2 that the value $a(\tau_{0})
/\tau_{0}$ decreases with the $\tau_{0}$.  Equation (12)
indicates that the mean energy gained by soft photons through first
scattering depends not only on the Compton parameter $y$, but also on
the optical depth $\tau_0$ of the corona. Only when $y\ll 1$, the mean
energy gained per scattering could be written as $\propto y$.
It is shown from Eq. (3) that the optical depth becomes smaller for the
higher electron temperature of the corona when the value of $y$ is fixed.
From Eqs. (12), (13) and Fig.2, we know that, for the same Compton
parameter $y$, there is  more energy gained by the soft photons through
single scattering for the corona with a smaller
optical depth $\tau_0$, i.e., a higher electron temperature $\Theta$.
This is the reason why the $D-y$
behaviour depicted in Fig. 1 varies with the electron temperature of the
corona.

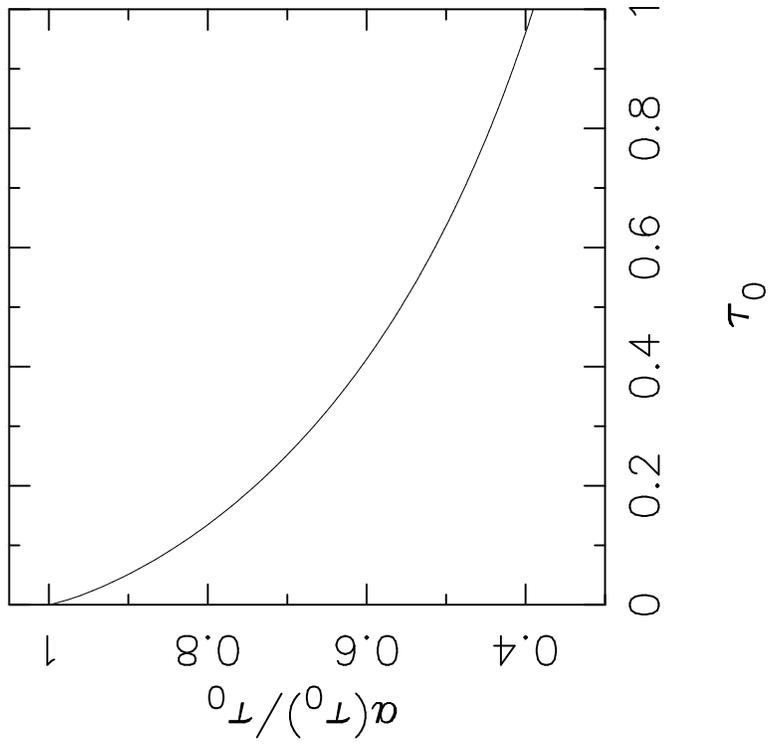
   \begin{figure}
      \vspace{8cm}
      \caption{The function $a(\tau_{0})/\tau_0$ vs. vertical optical depth $\tau_0$ 
of the corona.}
         \label{}
   \end{figure}

The relation $D\sim 2$ is required in the disk-corona system
(Haardt \& Maraschi 1991; Nakamura \& Osaki 1993). The Compton parameter
$y$ is about $0.6$ (Haardt \& Matt 1993). Our numerical results
show that $y\sim 0.35-0.4$ is more desirable to yield $D\sim 2$
corresponding to different electron temperatures $\Theta$.  
The difference may be
attributed to an ambiguity in measuring the optical depth of the system
with slab geometrical configuration. For the photon travels just in the
normal direction of the slab and the optical depth $\tau_0\ll 1$, the Compton
parameter $y$ well describes the situation. Nonetheless, most photons pass
the slab in the directions other than the vertical direction, these photons
will travel a longer distance than the vertical length before they leave
the corona since the disk is in slab geometry. The effective optical depth of these photons are
larger than $\tau_0$. Hence the Compton parameter $y$ defined by Eq. (3)
cannot describe the corona in slab configuration correctly.

It should be indicated that the present calculations are not effective
in the case that saturation of Compton scatterings is important.
The typical blackbody radiation temperature of the cold disk in
Active Galactic Nuclei is $\sim 5-50$ eV (Haardt \&
Maraschi 1993). Assuming the temperature of electron in the corona
$\Theta\sim 0.5$, the mean scattering number of 
the photon is $\sim 5$ before the Compton process saturates. For a lower
electron temperature $\Theta$, the photon suffers more scatterings before
it reaches saturation, for example, the mean scattering number is $\ge 20$ for
$\Theta\sim 0.1$. Hence, the saturation of the
Compton process in the disk should be taken into account at least in
some cases with relatively higher electron temperatures. In principle,
the recoil of the electron in the scattering and
Klein-Nishina electron scattering cross-section should be taken into account
in the study of the saturation of the Compton process for the mild
relativistic electrons treated here. For simplicity, we do not apply the
complicated formalisms including Klein-Nishina cross-section, instead,
we just repeat the calculations in the same way described at the beginning
of this section till the mean energy of the scattered photons exceeds
that of the electrons in the corona, and then let it simply be the
mean energy of the electrons. Yet, this numerical approach is still too
complicated to be used in constructing the disk-corona model. So, we propose
a semi-analytical approach to approximate the numerical results.

Suppose the number of photons which suffer at least $k$-fold scatterings
is $N_k$, the mean energy gained by the photons in the $k$-th scattering
is $N_k(1+A)^{k-1}A\epsilon_i$, where $\epsilon_i$ is the mean energy
of the input soft photons from the cold disk, $A=4\Theta+16\Theta^2$,
is the amplified factor. Thus, the fraction $D$ of the energy dissipated
in the hot corona to the input soft photon energy flux from the cold
disk is

\begin{equation}
D={1\over {N_{0}}}\sum\limits_{k=1}^{m}{N_{k}}(1+A)^{k-1}A,
\end{equation}
where $m$ is the scattering number of the photon before the Compton 
process saturates, $N_0$ is the number of input soft photons from
the cold disk. We know that the effective optical depth will be larger than
the vertical optical depth $\tau_0$. The fraction of the soft photons
suffering the first scattering is $a(\tau_0)$ given by Eq. (13). The first scattered
photon number ${N_1}={N_0}a(\tau_0)$. We further assume
${N_{k+1}}={N_k}b(\tau_0)$ for $k\geq 1$, where the coefficient
$b(\tau_0)$ is assumed to be a constant for the multiple scatterings 
except the first one for the photons. Then, $b(\tau_0)$
is only a function of the optical depth $\tau_0$.
Therefore, Eq. (14) could be rewritten as

\begin{equation}
D={{a(\tau_0)A[1-b(\tau_0)^{m}(1+A)^{m}]}\over {1-b(\tau_0)(1+A)}}.
\end{equation}
We take Eq. (15) as the 'seed' analytic form to fit the numerical solutions 
and find that 

\begin{equation}
D={{a(\tau_0)A[1-b(\tau_0)^{x}(1+A)^{x}]}\over {1-b(\tau_0)(1+A)}}.
\end{equation}
can well approximate the numerical results to an accuracy of $<2\%$ in 
the range of $D=0-5$ for the optically thin, hot corona. The two coefficients 
$a(\tau_0)$ and $b(\tau_0)$ in the formula are given by Eq. (13) and 

\begin{equation}
b(\tau_0)=(0.2302 \tau_{0}^{0.6956}+0.5102\tau_{0}){{a(\tau_0)}\over
{\tau_0}},
\end{equation}
respectively.  The mean effective scattering number $x$ for the photon
before it reaches saturation is

\begin{equation}
x=\log_{(1+A)} {\left({ {0.511\times10^{6}\Theta} \over {kT_{bb}} } \right)}
+0.8,
\end{equation}
where $0.8$ in the right side of the equation
is simply induced to fit the numerical results better, $kT_{bb}$ (in the
unit of eV) is the temperature of the blackbody radiation from the cold disk.
The comparisons between the semi-analytical approximations and the numerical
results of the problem are plotted in Figs. 3 and 4.

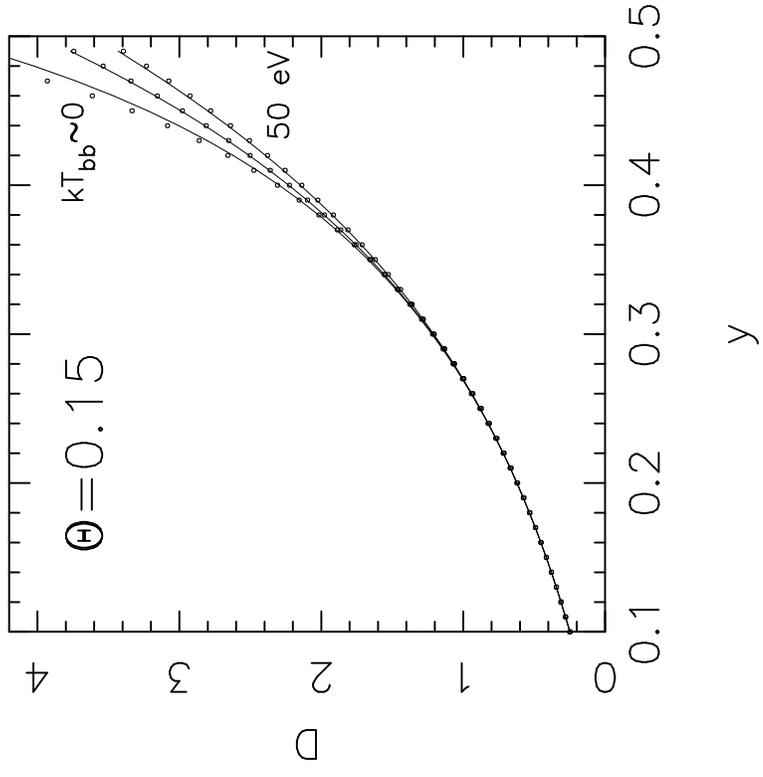
   \begin{figure}
      \vspace{8cm}
      \caption{Comparisons between the results of the semi-analytical
approximations and the numerical results (small circles) for the blackbody
radiation temperatures of the soft photons $kT_{bb}=0, 5$ and $50$ eV,
with respect to the electron temperature $\Theta=0.15$. }
         \label{}
   \end{figure}

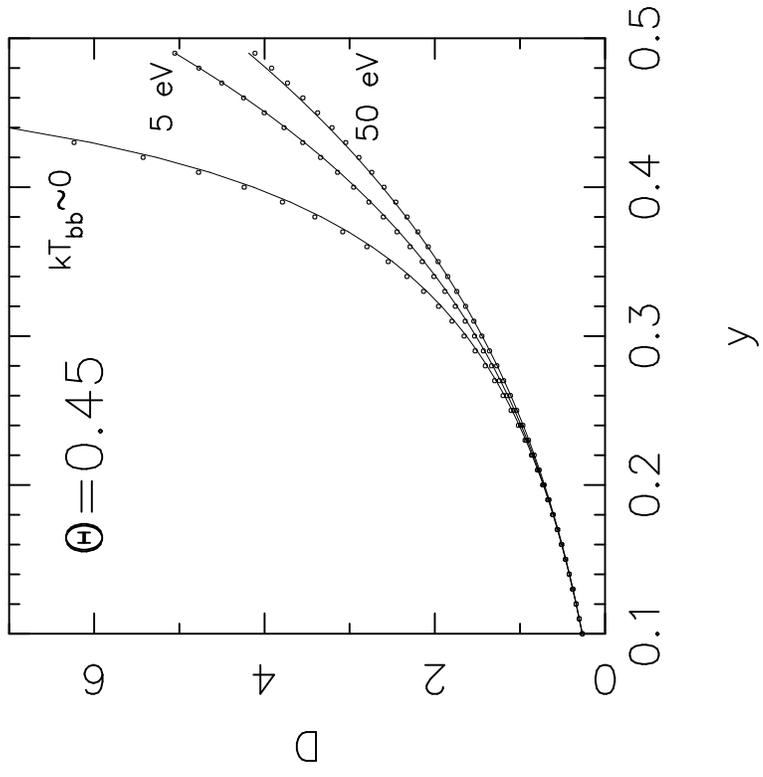
   \begin{figure}
      \vspace{8cm}
      \caption{Same as Fig. 3, but the electron temperature $\Theta=0.45$ instead. }
         \label{}
   \end{figure}

\section{Discussion}

We investigate the energy dissipations in the hot corona of the disk
by iterating the radiative transfer equation. A somewhat simplified 
method is employed to study the saturations of Compton processes. In Fig. 3
both the numerical results and semi-analytical approximations are plotted.
We find that the value of $D$ decreases with the energy of the soft photons
$kT_{bb}$ for the same electron temperature $\Theta$. The reason is that in
the limit of $y\gg 1$, all photons are in equilibrium with the electrons,  
$D\sim 0.511\times10^{6}\Theta /kT_{bb}$. Therefore, a larger $kT_{bb}$
for the input soft photons implies a smaller $D$ for the fixed electron
temperature $\Theta$. We note that the behaviours of $D-y$ do not
change significantly with the soft photons $kT_{bb}$ for relatively lower
electron temperatures (see Fig. 3 for $\Theta=0.15$). In this case, the
photon suffers more scatterings before it reaches saturation, which
indicates only a smaller fraction of input soft photons is finally in
energy equilibrium with the electrons in the corona.
The $D\sim 2$ is desired in two-phase model, which requires that the
Compton parameter $y$ to be $0.3-0.4$ with respect to the different
electron temperatures $\Theta$ and the blackbody radiation temperatures
$kT_{bb}$ of the photons from the cold disk. For relatively higher electron
temperature, there is a larger amplified factor $A$, and the Compton
process will soon saturate within several scatterings except
$k{T_{bb}}\sim 0$. Compared with the lower electron temperature case,
more soft photons are driven to equilibrium with the electrons, and then
no longer gain energy from hot electrons
through further scatterings. We suggest that Eq. (16) combining Eqs. (13),
(17) and (18) could be used as an approximate approach in the study of
the structure and the spectrum of the disk-corona systems.
Finally, we point that the pair production and 
annihilation in the hot corona should be taken into account when  the 
electron temperature in the corona exceeds $10^9$ K. This will be the scope 
of the further work. 

\begin{acknowledgements}
We thank the referee for his helpful suggestions.
The support from Pandeng Plan is gratefully acknowledged. XC thanks the
support from Shanghai Observatory, China Post-Doctoral Foundation and
NSFC.
\end{acknowledgements}

\end{document}